\documentclass[aps,prl,twocolumn,superscriptaddress,10pt]{revtex4-2}
\usepackage{siunitx}
\usepackage{amsfonts}
\usepackage{amsmath}
\usepackage{amssymb}
\usepackage{graphicx}
\usepackage{dcolumn}
\usepackage{mciteplus}
\usepackage{euscript}
\usepackage{multirow}
\usepackage{color}
\usepackage{amsmath}  
\usepackage{bm}
\usepackage{wasysym}    
\usepackage{float} 
\graphicspath{{./figs/}}

\begin{document}

\title{Topological Defect Mediated Helical Phase Reorientation by Uniaxial Stress}

\author{
  Tae-Hoon Kim}
  \thanks{T.K. and H.Z. contributed equally to this work.}
\affiliation{Ames National Laboratory, U.S. Department of Energy, Ames, Iowa 50011, USA}

\author{Haijun Zhao}
\thanks{T.K. and H.Z. contributed equally to this work.}
\affiliation{Ames National Laboratory, U.S. Department of Energy, Ames, Iowa 50011, USA}

\affiliation{
School of Physics, Southeast University, 211189 Nanjing, China.}
\author{ 
  Brandt A. Jensen}
  \affiliation{Ames National Laboratory, U.S. Department of Energy, Ames, Iowa 50011, USA}
\author{ 
Liqin Ke}
\thanks{Corresponding authors: liqinke@ameslab.gov and linzhou@ameslab.gov} 
\affiliation{Ames National Laboratory, U.S. Department of Energy, Ames, Iowa 50011, USA}
\author{ 
Lin Zhou 
}

\thanks{Corresponding authors: liqinke@ameslab.gov and linzhou@ameslab.gov} 

\affiliation{Ames National Laboratory, U.S. Department of Energy, Ames, Iowa 50011, USA}
\affiliation{Department of Materials Science and Engineering, Iowa State University, Ames, Iowa 50011, USA}

\date{\today}

\begin{abstract}
Strain engineering enables precise, energy-efficient control of nanoscale magnetism. However, unlike well-studied strain-dislocation interactions in mechanical deformation, the spatial evolution of strain-induced spin rearrangement remains poorly understood. Using \emph{in situ} Lorentz transmission electron microscopy, we manipulate and observe helical domain reorientation under quantitatively applied uniaxial tensile stress. Our findings reveal striking similarity to plastic deformation in metals, where the critical stress for propagation vector (\emph{\textbf{Q}}) reorientation depends on its angle with the stress direction. Magnetic defects mediate reorientation via “break-and-reconnect” or “dislocation gliding-annihilation” processes. Simulations confirm that strain-induced anisotropic Dzyaloshinskii-Moriya interaction may play a key role. These insights advance strain-driven magnetism and offer a promising route for energy-efficient magnetic nanophase control in next-generation information technology.
\end{abstract}
\maketitle

\begin{figure}
  \begin{center}
    \includegraphics[width=0.98\linewidth]{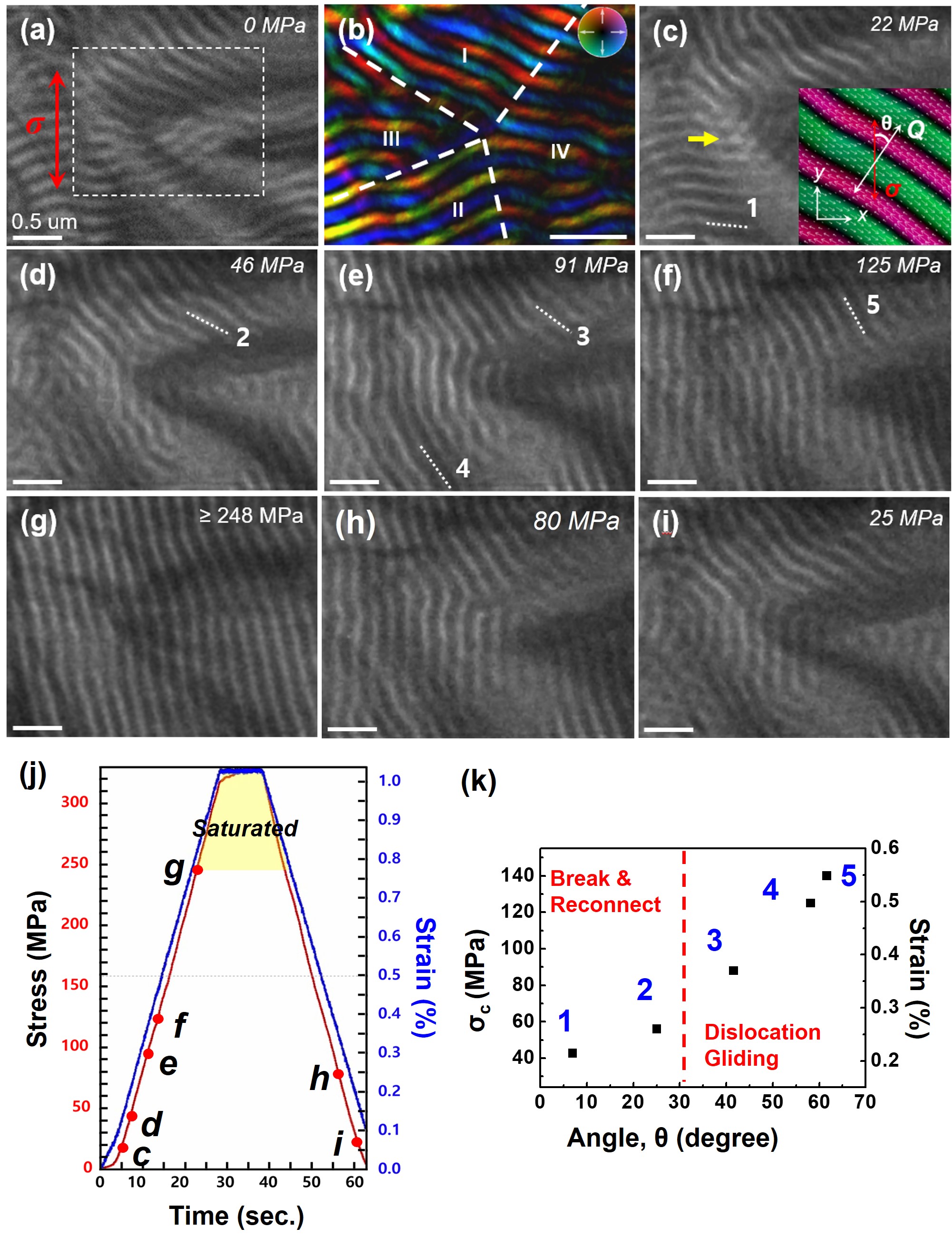}\\
  \end{center}
  \vspace{-15pt}
\caption{Sequence of domain rotation during loading and unloading. 
(a) Initial LTEM image showing domain distribution before stress application.
(b) In-plane magnetization map of helical phase revealing four domains from the dotted rectangle region in (a). The dashed  lines indicate the domain wall position and the color represents the in-plane magnetic induction direction.
(c)$-$(g) LTEM images illustrate the domain reorientation process. The sequence of reorientating domains is labeled as 1-5. 
Color inset and a yellow arrow in (c) illustrates the Q and stress ($\sigma$) directions and the quadruple junction, respectively. (h)$-$(i) LTEM 
images show recovering of the multidomain structure during the unloading process.
 (j) Temporal stress-strain curve obtained during
experiment. Loading conditions in (c)$-$(i) are labeled accordingly. (k) Plot of critical stress $\sigma_{c}$ (left) or strain (right) vs.
reorientation angle $\theta$, measured using domains 1$-$5 from (c)$-$(f).
}
\label{fig_ex1}
\vspace{-15pt}
\end{figure}

Chiral magnetism has attracted extensive research attention due to its
fundamental science and technological importance \cite{schoenherr2018topologicala,bogdanov2020emergence,kanazawa2017noncentrosymmetric,fert2017magnetic}.
In chiral magnet domains, the competition between exchange and
Dzyaloshinskii-Moriya
interaction (DMI) causes the spins to wind
periodically on a plane either perpendicularly (helix) or with a canting
angle (cone) along a specific direction, defined as a propagation vector
(\emph{\textbf{Q}})\cite{schoenherr2018topologicala,masell2020combinga} [Fig. \ref{fig_ex1}(c)]. Both intrinsic
crystal imperfections\cite{schoenherr2018topologicala} and external
stimuli\cite{masell2020combinga}, such as magnetic field, can change the
\emph{\textbf{Q}} vector\cite{schoenherr2018topologicala,masell2020combinga,rybakov2015newa,dussaux2016locala}. This change
may be accompanied by the appearance of a variety of nanometer scale
spin textures with topological charge, e.g., skyrmions or
merons\cite{schoenherr2018topologicala,dussaux2016locala}. Innovative application concepts, including
high-density memory and unconventional computing devices, have been
proposed based on control of the chiral magnet's complex spin configurations
by using magnetic field\cite{milde2013unwindinga,kim2020mechanismsa,kim2021kineticsa}, 
heat\cite{kim2021kineticsa,kong2013dynamicsa,yu2011roomtemperaturea}, electric current\cite{jiang2017directa,yu2012skyrmion,yu2017current},
microwave\cite{weiler2017helimagnona,mochizuki2012spinwavea}, and strain\cite{shibata2015largea,zhang2021straindrivena,lei2013straincontrolled,kong2023direct}.

Mechanical control of magnetic order is considered as a promising
ultra-low-energy consumption method at the nanometer
scale\cite{wang2019mechanical,bukharaev2018straintronics}. Additionally, integrating chiral magnetic
structures into current technology is likely to be based on thin-film
structures, usually fabricated by heteroepitaxial growth. The strain
introduced by the lattice mismatch between substrate and epilayer can
change the energy landscape of the system and affect the behavior of the
magnetic structure inside. As a result, a systematic understanding of
strain and chiral magnetic domain is critical for next generation
straintronics device design.

Experimentally, magnetization control via stress has been achieved in
magnetic thin film\cite{gopman2016strainassisted} and magnetostrictive
nanomagnets\cite{lei2013straincontrolled}. For chiral magnets, mechanical stress
can create, annihilate, or distort skyrmions\cite{sukhanov2019gianta,nii2015uniaxiala},
enhance skyrmion lattice stability\cite{shibata2015largea}, and induce
nontrivial spin textures\cite{paterson2020tensile}. These phenomena are
considered to be related to either strain induced magnetocrystalline
anisotropy\cite{sukhanov2019gianta,nii2015uniaxiala,plumer1982magnetoelastic} or modification of
DMI\cite{shibata2015largea,koretsune2015controla}. 

Simulation demonstrates that magnetic anisotropy
can not only adjust the spin spiral in helical phase but also induce
topological phase transitions of skyrmions\cite{sukhanov2019gianta}.
Alternatively, the modification of DMI was considered to be the reason
for strain induced elongation of skyrmions\cite{shibata2015largea}.    

In the
centrosymmetric material
La\textsubscript{0.67}Sr\textsubscript{0.33}MnO\textsubscript{3} 
where
DMI is absent, graded strain can induce DMI and leading to formation of
helical phase, and skyrmions\cite{zhang2021straindrivena}. 
In the noncentrosymmetric Mn$_{1-x}$Fe$_x$Ge  system, positive stress was found to reduce the DMI strength along the stress direction, modifying magnetic phases stability\cite{koretsune2015controla}.
Therefore, experimental investigation of the spatial evolution of domain structure and its relationship with precisely controlled loading magnitude is
 critical to corroborate the
theoretical models, but has not yet been demonstrated.

This study has achieved \emph{in situ} investigation of magnetic defects
mediated helical phase reorientation under quantitatively controlled
uniaxial tensile stress using Lorentz transmission electron microscopy
(LTEM). Our results revealed that the \emph{\textbf{Q}} vector direction
of the helical phase can be manipulated to the direction perpendicular
to the uniaxial stress. The strain-induced helical phase reorientation
initiates at the region with magnetic topological defects, in a manner
like plastic deformation in metal. The critical stress required to
induce helical phase reorientation is dependent on the angle between the
\textbf{\emph{Q}} and the direction of stress. Additionally, we
discovered two distinct reorientation mechanisms:
``break-and-reconnect'' and ``dislocation gliding-annihilation''
process, which are governed by the angle between \emph{\textbf{Q}} and
external stress as well. These results are supported by simulation and
analytical calculation based on strain induced anisotropic DMI theory.

A Co\textsubscript{8}Zn\textsubscript{8.5}Mn\textsubscript{3.5} (001)
thin plate    with {[}110{]} lateral orientation was fabricated from a large, multigrain crystal (methods are detailed in Supplementary Material \cite{Supplementary}) using focused ion beam and then transferred onto a push-to-pull (PTP) device.
 The thickness of the observed area is 190 nm. Tensile testing was performed along the [110] crystal direction by pushing the semi-circular part of the PTP device with a flat punch probe, which moved the mobile part
. Quantitative tensile load-displacement data were recorded with real-time imaging.
Details of the configuration of the device and the plate, as well as 
the geometric relationship between \emph{\textbf{Q}}, applied stress ($\sigma $), the angle
($\theta $) between $\sigma $ and \emph{\textbf{Q}}, and the \emph{x} and
\emph{y} coordinates directions [also illustrated in the inset of \ref{fig_ex1} (c)]
are described in the Supplementary Material \cite{Supplementary}.

\begin{figure}
  \begin{center}
    \includegraphics[width=0.98\linewidth]{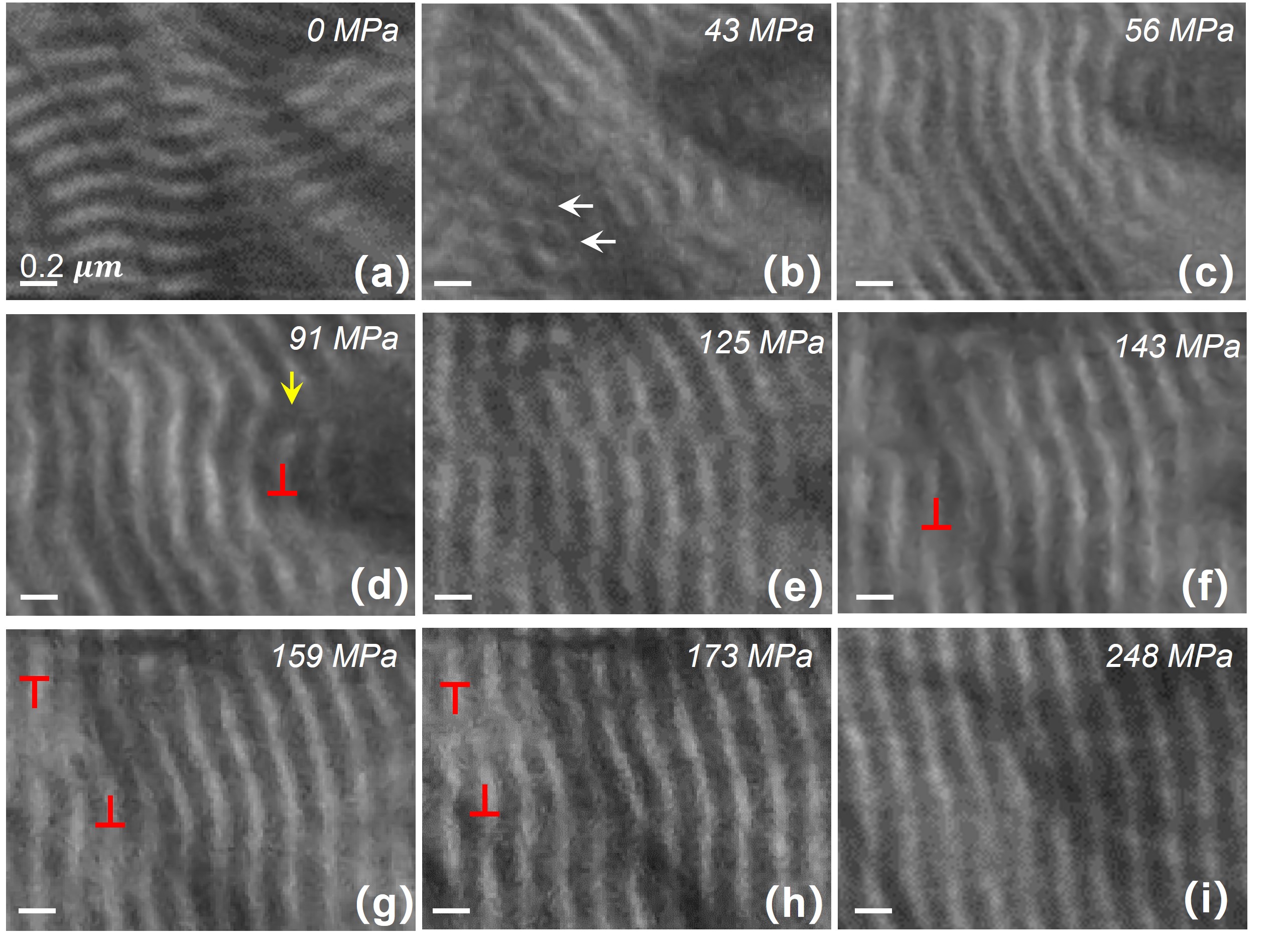}\\
  \end{center}
  \vspace{-15pt}
  \caption{Mechanisms of domain reorientation under varying stress loads:
  (a)$-$(c) Break and reconnect: the helix breaks at small $\theta$ [as shown by white arrows in (b)] and then reconnects (c).
  (d)$-$(f) Dislocation gliding: as stress increases from 91 MPa to 143 MPa, an edge dislocation (yellow arrow, red sign) glides from left to right. A bridge feature emerges transiently during this process (e).
  (g)$-$(i) Dislocation annihilation: two oppositely signed edge dislocations [red signs in (g)] merge as stress rises from 159 to 248 MPa. Scale bar: 0.2 $\mathrm{\mu}$m.
  }
  \vspace{-17pt}
  \label{fig_ex2}
\end{figure}
\emph{Helical phase reorientation by uniaxial strain---}
The Co\textsubscript{8}Zn\textsubscript{8.5}Mn\textsubscript{3.5} thin
plate exhibits a multidomain helical phase structure with a periodicity,
\(L_{D} \approx 135\) nm under zero field and stress, as shown by
LTEM image in Fig. \ref{fig_ex1}(a). The \emph{\textbf{Q}} of the helical phases lies
within the surface plane perpendicular to the white and black contrast
stripes. A helical phase may consist three types of domain walls based
on the angle between two \emph{\textbf{Q}} of the neighboring helices $\alpha$:
curved wall (type I, 
$\alpha \leqslant  85{^\circ}$)

zigzag disclination wall (type II,
$ 85{^\circ} \leqslant \alpha \leqslant 140{^\circ}$)
and dislocation wall (type III,
$\alpha \geqslant  140{^\circ}$ )\cite{schoenherr2018topologicala},
which are formed due to change of \emph{\textbf{Q}} direction.
Topological defects arise at the disclination and dislocation
walls\cite{schoenherr2018topologicala}, and their presence may affect the dynamics of
the magnetic phase under external perturbation, such as formation,
relaxation, and controllability by external stimuli\cite{masell2020combinga}.
An in-plane magnetization map from the LTEM image shows four dominant
domains, with domain walls marked by white dashed lines [Fig. \ref{fig_ex1}(b)]. The
I-IV and II-IV domain walls are topologically trivial curvature walls
(type I), whereas the I-III and II-III domain walls are topologically
nontrivial dislocation walls (type III) with topological defects inside.

Uniaxial stress applied to the
Co\textsubscript{8}Zn\textsubscript{8.5}Mn\textsubscript{3.5} 
laminar
aligns its helical phase's \emph{\textbf{Q}} perpendicular to the stress
direction. In our experiment, the tensile stress was applied by pulling
along the {[}110{]} crystal direction of the sample [$y$ direction as
indicated by red arrow in Fig. \ref{fig_ex1}(a)] with a displacement rate of 3 nm/s
using a PTP device, which corresponds to a strain rate of 0.04 \%/s.
The sample was held for 10 s at 1.12 \% of strain with 330 MPa of
stress, and then the applied stress was unloaded with the same rate used
for the initial stress loading. Sequence of domain rotation during the loading and
unloading process are shown in Figs. \ref{fig_ex1}(c)$-\ref{fig_ex1}$(i), whereas, corresponding temporal stress-strain
loading curves obtained during the \emph{in situ} mechanical experiment
are shown in Fig. \ref{fig_ex1}(j) \cite{Supplementary}.
 The helical phase reorientation started
near the quadruple junction at 22 MPa, as indicated by yellow arrows in
Fig. \ref{fig_ex1}(c), and then propagated into the other domains. Interestingly, the critical
stress that triggers helix reorientation
(\(\sigma\)\textsubscript{c}) depends on \emph{$\theta $ }. For example,
the helices at the bottom left [labeled as 1  in Fig. \ref{fig_ex1}(c)] with an initial
\emph{\textbf{Q}} almost parallel to the external stress direction
[$\theta\approx 0$ rotate at 46 MPa [Fig. \ref{fig_ex1}(d)]], whereas the
top domain [labeled as  2 in Fig. \ref{fig_ex1}(d)] maintains almost the initial
orientation. A continuous increasing of applied external stress triggers
the sequential rotation of helices from  2 to  5 in Figs. \ref{fig_ex1}(d)$-$\ref{fig_ex1}(f). The
measured \(\theta\) from domain  1 to  5 is $7^\circ$, 25.1$^\circ$, 41.6$^\circ$, 58.2$^\circ$, and 61.7$^\circ$, respectively. A plot demonstrating the relationship between
\emph{$\sigma$\textsubscript{c}} and \emph{$\theta $ } is shown in Fig. \ref{fig_ex1}(k). Data
points marked by  1$-$5 correspond to the helices  1 to  5 in Figs. \ref{fig_ex1}(c)$-$\ref{fig_ex1}(f). When
the external stress reached 248 MPa, a single domain helical phase with
its \emph{\textbf{Q}} almost perpendicular to the stress direction
(\(\theta \approx 90^\circ\)) was formed, as shown in Fig. \ref{fig_ex1}(g). Removing the
external stress recovers the initial multidomain helix structure, with
local domain reorientation following reverse sequence of the loading
process, as shown in Figs. \ref{fig_ex1}(h) and \ref{fig_ex1}(i). This \emph{in situ} tensile testing was
conducted under the elastic region, as shown by the linear strain-stress
relationship (Fig. S2 \cite{Supplementary}). Supplemental Material Movie S1 shows the whole reversible helical
phase reorientation process with real-time stress and strain measurement.

\emph{Defect-mediated \textbf{Q} reorientation mechanisms---}
Microscopically, the reorientation of helices proceeded through break-and-reconnect [Figs. \ref{fig_ex2}(a)$-$\ref{fig_ex2}(c)], dislocation gliding [Figs.
\ref{fig_ex2}(d)$-$\ref{fig_ex2}(f)], and dislocation annihilation [Fig.
\ref{fig_ex2}(g)$-$\ref{fig_ex2}(i)] mechanisms. When the helix's
\emph{\textbf{Q}} was almost parallel to the external stress direction 
($\theta\approx 0 $  ), it broke into several fragments as the external stress
reached 43 MPa in domain III, as shown by discontinuity of the black and
white stripes in the LTEM image in Fig. \ref{fig_ex2}(b). Subsequently, it was quickly
reconnected and formed slantingly aligned helices with
\(\theta \approx \pi/2\) when the stress reached 56 MPa [Fig. \ref{fig_ex2}(c)].

For helices marked by 3$-$5 in Figs. \ref{fig_ex2}(e) and \ref{fig_ex2}(f), the reorientation
process is mainly initiated by magnetic dislocation movement, which is
like edge dislocation (ED) gliding in real crystal. Figure \ref{fig_ex2}(d) shows an
ED in the helical phase, marked by a yellow arrow. The core of
dislocation is a topologically nontrivial meron\cite{schoenherr2018topologicala,masell2020combinga,yu2018transformation},
whose winding number (\(N_{W})\) is \(\pm 0.5\). With an increasing
applied external stress, the ED glides from right to left [Figs. \ref{fig_ex2}(d) to 
\ref{fig_ex2}(f)]. A zigzag bridgelike structure is captured during the dislocation
gliding. The bridgelike structure is an intermediate state formed due
to a change of \(N_{W}\) when the dislocation moves with a step of half the helical period 
\(L_{D}/2\) and at the same time changes its sign of \(N_{W}\). In
addition, oppositely charged ED will annihilate when they meet. As
illustrated in Fig. \ref{fig_ex2}(g)$-$\ref{fig_ex2}(i), an ED (indicated by a red ``\textsf{T}'' sign) moved
to the left-hand side and merged with another dislocation with the
opposite sign. This mechanism strikingly corroborates the similarity
between chiral magnets and cholesteric liquid crystals, where similar
rotational behavior of stripe domains was observed\cite{nose2010rotationala}.
The universally of our results highlights that the reorientation of the
helical magnetic structure is governed by the inherent
defects arising from the stripe structure.

\emph{Physical origin of $\mathbf{Q}$ and $\mathbf{\sigma}$ relationship---}
The key physical factor driving helix reorientation is potentially the strain-induced anisotropy in DMI strength\cite{shibata2015largea}. 

The helical phase emerges from the competition between exchange energy $E_{\text{ex}} = \int A (\nabla\mathbf{m})^2 dV$ and DMI energy $E_{\text{DM}} = \int \big(D_x \mathcal{L}_{yz}^{(x)}[\mathbf{m}] + D_y \mathcal{L}_{zx}^{(y)}[\mathbf{m}]\big) dV$, where $\mathbf{m} = (m_x, m_y, m_z)$ is the unit magnetization vector, $A$ is the exchange stiffness, $D_{x}$ ($D_{y}$) denotes DMI strength along $\hat{x}$ ($\hat{y}$), and $\mathcal{L}_{jk}^{(i)} = m_k \partial_i m_j - m_j \partial_i m_k$ defines the Lifshitz invariant.

Without loss of generality, let us assume that applying tensile strain ($\mathbf{\sigma}$) along the $\hat{y}$ direction reduces $D_y$  while leaving $D_x$ unaffected, thereby inducing anisotropy in the DMI strength.
This anisotropy is effectively captured by the parameter \(\eta(\sigma) = 1 - \frac{D_{y}(\sigma)}{D_{x}}\),
which increases from 0 to 1 as the strain  \(\sigma\) increases \cite{Supplementary}.

In the following section, we theoretically investigate the fundamental physics behind the two distinct helix reorientation processes by varying the effective strain parameter \(\eta\).
\begin{figure}[t]
  \begin{center}
    \includegraphics[width=0.98\linewidth]{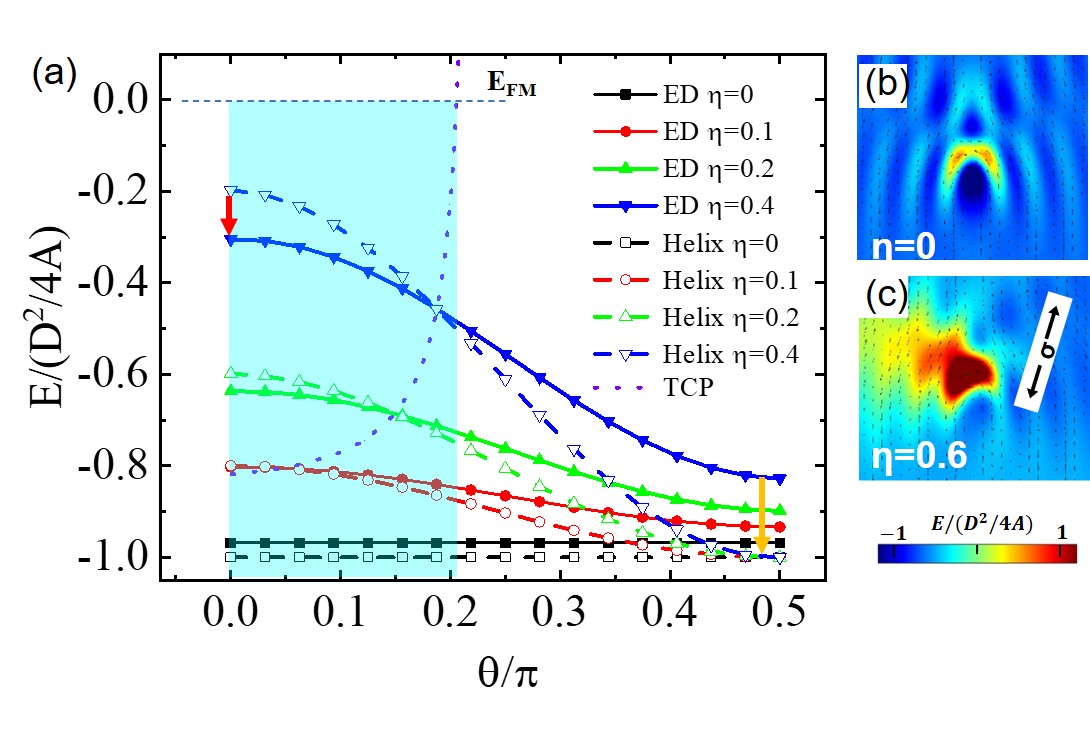}\\
  \end{center}
  \vspace{-20pt}
  \caption{Physical origin of the $\mathbf{Q}$ and $\mathbf{\sigma} $ relationship. (a) The energy of the
  helix phase with edge dislocation ($E$\textsubscript{ED}) and without edge
  dislocation ($E$\textsubscript{H}) as functions of $\theta $ at different
  $\eta $. The dotted line shows the trajectory of the crossover point
  (TCP) at which $E$\textsubscript{ED} = $E$\textsubscript{H}, when
  continuously increasing $\eta $. The TCP reaches the upper limit at $\theta \approx 0.2\pi$
  when $E$\textsubscript{ED} = $E$\textsubscript{H} approaches energy density
  of ferromagnetic state $E$\textsubscript{FM}=0. For the region above the TCP
  line, breaking the helix is favorable since $E$\textsubscript{ED}
  \textless $E$\textsubscript{H}, and this region lies in the small angle
  region ($\theta$\textless0.2$\pi$, the blue shaded region). For large $\theta $, strain
  promotes the annihilation of EDs since $E$\textsubscript{ED}\textgreater $E$\textsubscript{H}. The red and yellow arrows show the
  energy release of ED generation or annihilation, at small or large $\theta $,
  respectively. (b) and (c) Plot of energy distribution around an
  individual edge dislocation core in the absence of stress, or in the
  presence of stress ($\theta=13\pi/32\approx 73^\circ $, $\eta =0.6$), respectively.
  The imbalanced energy around the edge dislocation under external strain
  [as shown in  (c)] causes it to glide.}
  \label{fig_Eed}
  \vspace{-15pt}
\end{figure}

\emph{Physics behind break-and-reconnect reorientation:}
Based on the understanding that helix disconnections and reconnections lead to the creation or elimination of ED pairs, we numerically evaluate the energy of an isolated ED to reveal the fundamental physics behind the helix break-and-reconnect reorientation process. 
An ED can be stabilized by slightly misaligning the $\mathbf{Q}$ vectors of two adjacent helical domains by $\alpha = 0.9^\circ$. Subsequently, we determine the ED energy $E$\textsubscript{ED} by averaging the energy density over a $1.5L_{D} \times 1.5L_{D}$ region centered around the ED's core and compare this value to the energy $E$\textsubscript{H} of an unperturbed helix configuration.

Figure \ref{fig_Eed}(a) compares $E$\textsubscript{ED} and $E$\textsubscript{H} as functions of $\theta$ for various $\eta$. At finite $\eta$, both energies increase as  $\theta$ decreases, but $E$\textsubscript{ED} rises at a slower rate. This results in ED configurations becoming more stable than the helix for smaller $\theta$ ($E_{\text{ED}}<E_{\text{H}}$) and less stable for larger  $\theta$ ($E_{\text{ED}}>E_{\text{H}}$). The crossover angle $\theta_{\mathrm{c}}$ where \(E_{\text{ED}} = E_{\text{H}}\) increases with $\eta$ up to a maximum value of approximately $\theta_{\mathrm{c}}^{\max} = 36^\circ$, at which point $E$\textsubscript{ED} converges to the ferromagnetic state energy $E_{\text{FM}}$. For reorientation angle above $\theta_{\mathrm{c}}^{\max}$, 
$E$\textsubscript{H} remains lower than $E$\textsubscript{ED}, stabilizing the pure helix configuration and inhibiting helix breaking.

These findings explain our experimental observations of the break-and-reconnect process. For small-angle helices, breaking occurs when  $E$\textsubscript{ED}$<E$\textsubscript{H}, generating small helix fragments. As these fragments rotate and their angles increase, $E$\textsubscript{ED} eventually exceeds $E$\textsubscript{H}, favoring reconnection to minimize fragmented elements. Notably, the numerically derived $\theta_{\mathrm{c}}^{\max}=36^\circ$  quantitatively aligns with our experimental observations, acting as a clear boundary between distinct reorientation mechanisms.

\begin{figure}
  \begin{center}
    \includegraphics[width=0.98\linewidth]{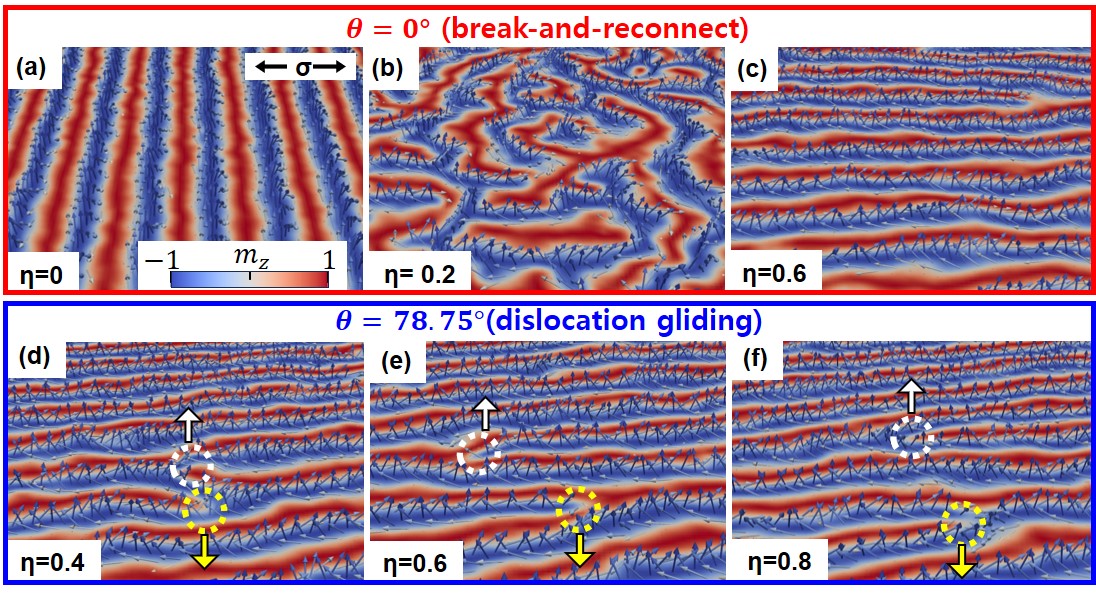}\\
  \end{center}
  \vspace{-15pt}
  \caption{ Micromagnetic simulations of the spin reorientation
  inside the helical phase under uniaxial stress. (a)$-$(c) domain structure
  evolution of a helix with \(\theta = 0\) with increasing \(\eta\) from 0
  to 0.6. A single-$Q$ helix (a) breaks into a disorder mixture of
  short helices and skyrmions (b), followed by reconnection,
  forming a single-$Q$ helix with \(\theta \approx \pi\text{/}2\ \)in
  (c). (d)$-$(f) For large angle \(\theta = 7\pi\text{/}16\),
  the reorientation follows the edge dislocation mediated process: the
  anisotropic DMI pushes the preexisting positive and negative dislocation
  gliding in opposite directions with increasing \(\eta\ \)from 0.4 to
  0.8.
  }
  \label{fig_MMS}
  \vspace{-15pt}
\end{figure}
\emph{Physics behind dislocation gliding process:}
For larger
\(\theta\), the reorientation of the helix through the creation and annihilation of edge dislocations is energetically unfavorable. However, gliding of individual EDs persists as a reorientation mechanism. 
Figures \ref{fig_Eed}(b) and \ref{fig_Eed}(c) illustrate the energy distribution around an ED core,  with and without external stress applied \((\theta \approx 73{^\circ}\), \(\eta = 0.6)\). 
Without stress, the energy distribution is nearly symmetric along the \(\mathbf{Q}\) direction, resulting in zero force on the ED. Under stress, however, the distribution becomes asymmetric, inducing a nonzero force that activates ED gliding. This gliding slightly rotates neighboring helices, partially releasing accumulated energy.

\emph{Helix reorientation in micromagnetic:}
To better understand the
spatial evolution of the spin rearrangement under uniaxial stress, we
performed a micromagnetic simulation. Figure \ref{fig_MMS} shows the simulated
results of the spin reorientation process of a helix with
\(\theta\  = 0\) [Figs. \ref{fig_MMS}(a)$-$\ref{fig_MMS}(c)] and \(\theta = 7\pi\text{/}16\) [Figs. \ref{fig_MMS}
(d)$-\ref{fig_MMS}$(e)] under increasing \(\eta.\) For \(\theta = 0\), similar to our
experiment, the reorientation follows the break and reconnection
process: a single-$Q$ helix [Fig. \ref{fig_MMS}(a)] breaks into a disordered mixture of
short helices and skyrmions [Fig. \ref{fig_MMS}(b)], followed by reconnection, forming
single-$Q$ helix with \(\theta \approx \pi\text{/}2\) [Fig. \ref{fig_MMS}(c)]. For
\(\theta = 7\pi\text{/}16\), the reorientation follows the ED mediated
process: The anisotropic DMI pushes the preexisting positive and
negative ED gliding in opposite directions [see Fig. \ref{fig_MMS}(d)$-$\ref{fig_MMS}(f)] until they
disappear at a boundary or merge with another one with opposite Burger's
vector.

In summary, we have investigated the defect-mediated reorientation of
the helical phase through \emph{in situ} LTEM tensile experiments. Our
results demonstrate that the helical phase's \emph{\textbf{Q}} can be
reoriented perpendicular to the stress direction in the absence of an
external magnetic field. The reorientation of the helical phase is
initiated at magnetic topological defects and progresses through
defect-mediated mechanisms, specifically the break-and-reconnect or
dislocation gliding-annihilation process. Furthermore, we have
established that the critical stress required to initiate the
reorientation of the helical phase is strongly correlated to the angle
between the stress and the \emph{\textbf{Q}} vector of the helices. Our
simulations and calculations provide compelling evidence that strain
induced anisotropic DMI is likely responsible for the observed reorientation of
the helical phase. Our findings provide valuable mechanistic insights into the strain engineering of magnetic states, creating a foundation for future in-depth investigations into other topological magnetic states, such as skyrmions, and thus have the potential to expand the applications of spintronics and straintronics.

\begin{acknowledgments}
This work is supported in part by Laboratory Directed Research and Development funds through Ames National Laboratory (T.K., H.Z., L.Z.). L.K. was supported by the U.S. Department of Energy, Office of Science, Office of Basic Energy Sciences, Materials Sciences and Engineering Division, Early Career Research Program following conception and initial work supported by LDRD. Ames National Laboratory is operated for the U.S. Department of Energy by Iowa State University under Contract No. DE-AC02-07CH11358. 
H.Z. thanks the computational resources from the Southeast University Campus-Wide Computing Platform.
H.Z. is also supported in part by National Natural Science Foundation of China (No. 11704067). All TEM and related work were performed using instruments in the Sensitive Instrument Facility in Ames National Lab. Current affiliation of T.K.: Department of Materials Science and Engineering, Chonnam National University, Gwangju 61186, Republic of Korea.
\end{acknowledgments}


\begin{thebibliography}{40}%
\makeatletter
\providecommand \@ifxundefined [1]{%
 \@ifx{#1\undefined}
}%
\providecommand \@ifnum [1]{%
 \ifnum #1\expandafter \@firstoftwo
 \else \expandafter \@secondoftwo
 \fi
}%
\providecommand \@ifx [1]{%
 \ifx #1\expandafter \@firstoftwo
 \else \expandafter \@secondoftwo
 \fi
}%
\providecommand \natexlab [1]{#1}%
\providecommand \enquote  [1]{``#1''}%
\providecommand \bibnamefont  [1]{#1}%
\providecommand \bibfnamefont [1]{#1}%
\providecommand \citenamefont [1]{#1}%
\providecommand \href@noop [0]{\@secondoftwo}%
\providecommand \href [0]{\begingroup \@sanitize@url \@href}%
\providecommand \@href[1]{\@@startlink{#1}\@@href}%
\providecommand \@@href[1]{\endgroup#1\@@endlink}%
\providecommand \@sanitize@url [0]{\catcode `\\12\catcode `\$12\catcode `\&12\catcode `\#12\catcode `\^12\catcode `\_12\catcode `\%12\relax}%
\providecommand \@@startlink[1]{}%
\providecommand \@@endlink[0]{}%
\providecommand \url  [0]{\begingroup\@sanitize@url \@url }%
\providecommand \@url [1]{\endgroup\@href {#1}{\urlprefix }}%
\providecommand \urlprefix  [0]{URL }%
\providecommand \Eprint [0]{\href }%
\providecommand \doibase [0]{http://dx.doi.org/}%
\providecommand \selectlanguage [0]{\@gobble}%
\providecommand \bibinfo  [0]{\@secondoftwo}%
\providecommand \bibfield  [0]{\@secondoftwo}%
\providecommand \translation [1]{[#1]}%
\providecommand \BibitemOpen [0]{}%
\providecommand \bibitemStop [0]{}%
\providecommand \bibitemNoStop [0]{.\EOS\space}%
\providecommand \EOS [0]{\spacefactor3000\relax}%
\providecommand \BibitemShut  [1]{\csname bibitem#1\endcsname}%
\let\auto@bib@innerbib\@empty
%</preamble>
\bibitem [{\citenamefont {Schoenherr}\ \emph {et~al.}(2018)\citenamefont {Schoenherr}, \citenamefont {M{\"u}ller}, \citenamefont {K{\"o}hler}, \citenamefont {Rosch}, \citenamefont {Kanazawa}, \citenamefont {Tokura}, \citenamefont {Garst},\ and\ \citenamefont {Meier}}]{schoenherr2018topologicala}%
  \BibitemOpen
  \bibfield  {author} {\bibinfo {author} {\bibfnamefont {P.}~\bibnamefont {Schoenherr}}, \bibinfo {author} {\bibfnamefont {J.}~\bibnamefont {M{\"u}ller}}, \bibinfo {author} {\bibfnamefont {L.}~\bibnamefont {K{\"o}hler}}, \bibinfo {author} {\bibfnamefont {A.}~\bibnamefont {Rosch}}, \bibinfo {author} {\bibfnamefont {N.}~\bibnamefont {Kanazawa}}, \bibinfo {author} {\bibfnamefont {Y.}~\bibnamefont {Tokura}}, \bibinfo {author} {\bibfnamefont {M.}~\bibnamefont {Garst}}, \ and\ \bibinfo {author} {\bibfnamefont {D.}~\bibnamefont {Meier}},\ }\href {\doibase 10.1038/s41567-018-0056-5} {\bibfield  {journal} {\bibinfo  {journal} {Nature Phys}\ }\textbf {\bibinfo {volume} {14}},\ \bibinfo {pages} {465} (\bibinfo {year} {2018})}\BibitemShut {NoStop}%
\bibitem [{\citenamefont {Bogdanov}\ and\ \citenamefont {Panagopoulos}(2020)}]{bogdanov2020emergence}%
  \BibitemOpen
  \bibfield  {author} {\bibinfo {author} {\bibfnamefont {A.~N.}\ \bibnamefont {Bogdanov}}\ and\ \bibinfo {author} {\bibfnamefont {C.}~\bibnamefont {Panagopoulos}},\ }\href {\doibase 10.1063/PT.3.4431} {\bibfield  {journal} {\bibinfo  {journal} {Physics Today}\ }\textbf {\bibinfo {volume} {73}},\ \bibinfo {pages} {44} (\bibinfo {year} {2020})}\BibitemShut {NoStop}%
\bibitem [{\citenamefont {Kanazawa}\ \emph {et~al.}(2017)\citenamefont {Kanazawa}, \citenamefont {Seki},\ and\ \citenamefont {Tokura}}]{kanazawa2017noncentrosymmetric}%
  \BibitemOpen
  \bibfield  {author} {\bibinfo {author} {\bibfnamefont {N.}~\bibnamefont {Kanazawa}}, \bibinfo {author} {\bibfnamefont {S.}~\bibnamefont {Seki}}, \ and\ \bibinfo {author} {\bibfnamefont {Y.}~\bibnamefont {Tokura}},\ }\href {\doibase 10.1002/adma.201603227} {\bibfield  {journal} {\bibinfo  {journal} {Advanced Materials}\ }\textbf {\bibinfo {volume} {29}},\ \bibinfo {pages} {1603227} (\bibinfo {year} {2017})}\BibitemShut {NoStop}%
\bibitem [{\citenamefont {Fert}\ \emph {et~al.}(2017)\citenamefont {Fert}, \citenamefont {Reyren},\ and\ \citenamefont {Cros}}]{fert2017magnetic}%
  \BibitemOpen
  \bibfield  {author} {\bibinfo {author} {\bibfnamefont {A.}~\bibnamefont {Fert}}, \bibinfo {author} {\bibfnamefont {N.}~\bibnamefont {Reyren}}, \ and\ \bibinfo {author} {\bibfnamefont {V.}~\bibnamefont {Cros}},\ }\href {\doibase 10.1038/natrevmats.2017.31} {\bibfield  {journal} {\bibinfo  {journal} {Nat Rev Mater}\ }\textbf {\bibinfo {volume} {2}},\ \bibinfo {pages} {17031} (\bibinfo {year} {2017})}\BibitemShut {NoStop}%
\bibitem [{\citenamefont {Masell}\ \emph {et~al.}(2020)\citenamefont {Masell}, \citenamefont {Yu}, \citenamefont {Kanazawa}, \citenamefont {Tokura},\ and\ \citenamefont {Nagaosa}}]{masell2020combinga}%
  \BibitemOpen
  \bibfield  {author} {\bibinfo {author} {\bibfnamefont {J.}~\bibnamefont {Masell}}, \bibinfo {author} {\bibfnamefont {X.}~\bibnamefont {Yu}}, \bibinfo {author} {\bibfnamefont {N.}~\bibnamefont {Kanazawa}}, \bibinfo {author} {\bibfnamefont {Y.}~\bibnamefont {Tokura}}, \ and\ \bibinfo {author} {\bibfnamefont {N.}~\bibnamefont {Nagaosa}},\ }\href {\doibase 10.1103/PhysRevB.102.180402} {\bibfield  {journal} {\bibinfo  {journal} {Phys. Rev. B}\ }\textbf {\bibinfo {volume} {102}},\ \bibinfo {pages} {180402} (\bibinfo {year} {2020})}\BibitemShut {NoStop}%
\bibitem [{\citenamefont {Rybakov}\ \emph {et~al.}(2015)\citenamefont {Rybakov}, \citenamefont {Borisov}, \citenamefont {Bl{\"u}gel},\ and\ \citenamefont {Kiselev}}]{rybakov2015newa}%
  \BibitemOpen
  \bibfield  {author} {\bibinfo {author} {\bibfnamefont {F.~N.}\ \bibnamefont {Rybakov}}, \bibinfo {author} {\bibfnamefont {A.~B.}\ \bibnamefont {Borisov}}, \bibinfo {author} {\bibfnamefont {S.}~\bibnamefont {Bl{\"u}gel}}, \ and\ \bibinfo {author} {\bibfnamefont {N.~S.}\ \bibnamefont {Kiselev}},\ }\href {\doibase 10.1103/PhysRevLett.115.117201} {\bibfield  {journal} {\bibinfo  {journal} {Phys. Rev. Lett.}\ }\textbf {\bibinfo {volume} {115}},\ \bibinfo {pages} {117201} (\bibinfo {year} {2015})}\BibitemShut {NoStop}%
\bibitem [{\citenamefont {Dussaux}\ \emph {et~al.}(2016)\citenamefont {Dussaux}, \citenamefont {Schoenherr}, \citenamefont {Koumpouras}, \citenamefont {Chico}, \citenamefont {Chang}, \citenamefont {Lorenzelli}, \citenamefont {Kanazawa}, \citenamefont {Tokura}, \citenamefont {Garst}, \citenamefont {Bergman}, \citenamefont {Degen},\ and\ \citenamefont {Meier}}]{dussaux2016locala}%
  \BibitemOpen
  \bibfield  {author} {\bibinfo {author} {\bibfnamefont {A.}~\bibnamefont {Dussaux}}, \bibinfo {author} {\bibfnamefont {P.}~\bibnamefont {Schoenherr}}, \bibinfo {author} {\bibfnamefont {K.}~\bibnamefont {Koumpouras}}, \bibinfo {author} {\bibfnamefont {J.}~\bibnamefont {Chico}}, \bibinfo {author} {\bibfnamefont {K.}~\bibnamefont {Chang}}, \bibinfo {author} {\bibfnamefont {L.}~\bibnamefont {Lorenzelli}}, \bibinfo {author} {\bibfnamefont {N.}~\bibnamefont {Kanazawa}}, \bibinfo {author} {\bibfnamefont {Y.}~\bibnamefont {Tokura}}, \bibinfo {author} {\bibfnamefont {M.}~\bibnamefont {Garst}}, \bibinfo {author} {\bibfnamefont {A.}~\bibnamefont {Bergman}}, \bibinfo {author} {\bibfnamefont {C.~L.}\ \bibnamefont {Degen}}, \ and\ \bibinfo {author} {\bibfnamefont {D.}~\bibnamefont {Meier}},\ }\href {\doibase 10.1038/ncomms12430} {\bibfield  {journal} {\bibinfo  {journal} {Nat Commun}\ }\textbf {\bibinfo {volume} {7}},\ \bibinfo {pages} {12430} (\bibinfo {year} {2016})}\BibitemShut {NoStop}%
\bibitem [{\citenamefont {Milde}\ \emph {et~al.}(2013)\citenamefont {Milde}, \citenamefont {K{\"o}hler}, \citenamefont {Seidel}, \citenamefont {Eng}, \citenamefont {Bauer}, \citenamefont {Chacon}, \citenamefont {Kindervater}, \citenamefont {M{\"u}hlbauer}, \citenamefont {Pfleiderer}, \citenamefont {Buhrandt}, \citenamefont {Sch{\"u}tte},\ and\ \citenamefont {Rosch}}]{milde2013unwindinga}%
  \BibitemOpen
  \bibfield  {author} {\bibinfo {author} {\bibfnamefont {P.}~\bibnamefont {Milde}}, \bibinfo {author} {\bibfnamefont {D.}~\bibnamefont {K{\"o}hler}}, \bibinfo {author} {\bibfnamefont {J.}~\bibnamefont {Seidel}}, \bibinfo {author} {\bibfnamefont {L.~M.}\ \bibnamefont {Eng}}, \bibinfo {author} {\bibfnamefont {A.}~\bibnamefont {Bauer}}, \bibinfo {author} {\bibfnamefont {A.}~\bibnamefont {Chacon}}, \bibinfo {author} {\bibfnamefont {J.}~\bibnamefont {Kindervater}}, \bibinfo {author} {\bibfnamefont {S.}~\bibnamefont {M{\"u}hlbauer}}, \bibinfo {author} {\bibfnamefont {C.}~\bibnamefont {Pfleiderer}}, \bibinfo {author} {\bibfnamefont {S.}~\bibnamefont {Buhrandt}}, \bibinfo {author} {\bibfnamefont {C.}~\bibnamefont {Sch{\"u}tte}}, \ and\ \bibinfo {author} {\bibfnamefont {A.}~\bibnamefont {Rosch}},\ }\href {\doibase 10.1126/science.1234657} {\bibfield  {journal} {\bibinfo  {journal} {Science}\ }\textbf {\bibinfo {volume} {340}},\ \bibinfo {pages} {1076} (\bibinfo {year} {2013})}\BibitemShut {NoStop}%
\bibitem [{\citenamefont {Kim}\ \emph {et~al.}(2020)\citenamefont {Kim}, \citenamefont {Zhao}, \citenamefont {Xu}, \citenamefont {Jensen}, \citenamefont {King}, \citenamefont {Kramer}, \citenamefont {Nan}, \citenamefont {Ke},\ and\ \citenamefont {Zhou}}]{kim2020mechanismsa}%
  \BibitemOpen
  \bibfield  {author} {\bibinfo {author} {\bibfnamefont {T.-H.}\ \bibnamefont {Kim}}, \bibinfo {author} {\bibfnamefont {H.}~\bibnamefont {Zhao}}, \bibinfo {author} {\bibfnamefont {B.}~\bibnamefont {Xu}}, \bibinfo {author} {\bibfnamefont {B.~A.}\ \bibnamefont {Jensen}}, \bibinfo {author} {\bibfnamefont {A.~H.}\ \bibnamefont {King}}, \bibinfo {author} {\bibfnamefont {M.~J.}\ \bibnamefont {Kramer}}, \bibinfo {author} {\bibfnamefont {C.}~\bibnamefont {Nan}}, \bibinfo {author} {\bibfnamefont {L.}~\bibnamefont {Ke}}, \ and\ \bibinfo {author} {\bibfnamefont {L.}~\bibnamefont {Zhou}},\ }\href {\doibase 10.1021/acs.nanolett.0c00080} {\bibfield  {journal} {\bibinfo  {journal} {Nano Lett.}\ }\textbf {\bibinfo {volume} {20}},\ \bibinfo {pages} {4731} (\bibinfo {year} {2020})}\BibitemShut {NoStop}%
\bibitem [{\citenamefont {Kim}\ \emph {et~al.}(2021)\citenamefont {Kim}, \citenamefont {Zhao}, \citenamefont {Ong}, \citenamefont {Jensen}, \citenamefont {Cui}, \citenamefont {King}, \citenamefont {Ke},\ and\ \citenamefont {Zhou}}]{kim2021kineticsa}%
  \BibitemOpen
  \bibfield  {author} {\bibinfo {author} {\bibfnamefont {T.-H.}\ \bibnamefont {Kim}}, \bibinfo {author} {\bibfnamefont {H.}~\bibnamefont {Zhao}}, \bibinfo {author} {\bibfnamefont {P.-V.}\ \bibnamefont {Ong}}, \bibinfo {author} {\bibfnamefont {B.~A.}\ \bibnamefont {Jensen}}, \bibinfo {author} {\bibfnamefont {B.}~\bibnamefont {Cui}}, \bibinfo {author} {\bibfnamefont {A.~H.}\ \bibnamefont {King}}, \bibinfo {author} {\bibfnamefont {L.}~\bibnamefont {Ke}}, \ and\ \bibinfo {author} {\bibfnamefont {L.}~\bibnamefont {Zhou}},\ }\href {\doibase 10.1021/acs.nanolett.1c00923} {\bibfield  {journal} {\bibinfo  {journal} {Nano Lett.}\ }\textbf {\bibinfo {volume} {21}},\ \bibinfo {pages} {5547} (\bibinfo {year} {2021})}\BibitemShut {NoStop}%
\bibitem [{\citenamefont {Kong}\ and\ \citenamefont {Zang}(2013)}]{kong2013dynamicsa}%
  \BibitemOpen
  \bibfield  {author} {\bibinfo {author} {\bibfnamefont {L.}~\bibnamefont {Kong}}\ and\ \bibinfo {author} {\bibfnamefont {J.}~\bibnamefont {Zang}},\ }\href {\doibase 10.1103/PhysRevLett.111.067203} {\bibfield  {journal} {\bibinfo  {journal} {Phys. Rev. Lett.}\ }\textbf {\bibinfo {volume} {111}},\ \bibinfo {pages} {067203} (\bibinfo {year} {2013})}\BibitemShut {NoStop}%
\bibitem [{\citenamefont {Yu}\ \emph {et~al.}(2011)\citenamefont {Yu}, \citenamefont {Kanazawa}, \citenamefont {Onose}, \citenamefont {Kimoto}, \citenamefont {Zhang}, \citenamefont {Ishiwata}, \citenamefont {Matsui},\ and\ \citenamefont {Tokura}}]{yu2011roomtemperaturea}%
  \BibitemOpen
  \bibfield  {author} {\bibinfo {author} {\bibfnamefont {X.~Z.}\ \bibnamefont {Yu}}, \bibinfo {author} {\bibfnamefont {N.}~\bibnamefont {Kanazawa}}, \bibinfo {author} {\bibfnamefont {Y.}~\bibnamefont {Onose}}, \bibinfo {author} {\bibfnamefont {K.}~\bibnamefont {Kimoto}}, \bibinfo {author} {\bibfnamefont {W.~Z.}\ \bibnamefont {Zhang}}, \bibinfo {author} {\bibfnamefont {S.}~\bibnamefont {Ishiwata}}, \bibinfo {author} {\bibfnamefont {Y.}~\bibnamefont {Matsui}}, \ and\ \bibinfo {author} {\bibfnamefont {Y.}~\bibnamefont {Tokura}},\ }\href {\doibase 10.1038/nmat2916} {\bibfield  {journal} {\bibinfo  {journal} {Nature Mater}\ }\textbf {\bibinfo {volume} {10}},\ \bibinfo {pages} {106} (\bibinfo {year} {2011})}\BibitemShut {NoStop}%
\bibitem [{\citenamefont {Jiang}\ \emph {et~al.}(2017)\citenamefont {Jiang}, \citenamefont {Zhang}, \citenamefont {Yu}, \citenamefont {Zhang}, \citenamefont {Wang}, \citenamefont {Benjamin~Jungfleisch}, \citenamefont {Pearson}, \citenamefont {Cheng}, \citenamefont {Heinonen}, \citenamefont {Wang}, \citenamefont {Zhou}, \citenamefont {Hoffmann},\ and\ \citenamefont {{te~Velthuis}}}]{jiang2017directa}%
  \BibitemOpen
  \bibfield  {author} {\bibinfo {author} {\bibfnamefont {W.}~\bibnamefont {Jiang}}, \bibinfo {author} {\bibfnamefont {X.}~\bibnamefont {Zhang}}, \bibinfo {author} {\bibfnamefont {G.}~\bibnamefont {Yu}}, \bibinfo {author} {\bibfnamefont {W.}~\bibnamefont {Zhang}}, \bibinfo {author} {\bibfnamefont {X.}~\bibnamefont {Wang}}, \bibinfo {author} {\bibfnamefont {M.}~\bibnamefont {Benjamin~Jungfleisch}}, \bibinfo {author} {\bibfnamefont {J.~E.}\ \bibnamefont {Pearson}}, \bibinfo {author} {\bibfnamefont {X.}~\bibnamefont {Cheng}}, \bibinfo {author} {\bibfnamefont {O.}~\bibnamefont {Heinonen}}, \bibinfo {author} {\bibfnamefont {K.~L.}\ \bibnamefont {Wang}}, \bibinfo {author} {\bibfnamefont {Y.}~\bibnamefont {Zhou}}, \bibinfo {author} {\bibfnamefont {A.}~\bibnamefont {Hoffmann}}, \ and\ \bibinfo {author} {\bibfnamefont {S.~G.~E.}\ \bibnamefont {{te~Velthuis}}},\ }\href {\doibase 10.1038/nphys3883} {\bibfield  {journal} {\bibinfo  {journal} {Nature Phys}\ }\textbf {\bibinfo {volume} {13}},\ \bibinfo {pages} {162} (\bibinfo {year} {2017})}\BibitemShut {NoStop}%
\bibitem [{\citenamefont {Yu}\ \emph {et~al.}(2012)\citenamefont {Yu}, \citenamefont {Kanazawa}, \citenamefont {Zhang}, \citenamefont {Nagai}, \citenamefont {Hara}, \citenamefont {Kimoto}, \citenamefont {Matsui}, \citenamefont {Onose},\ and\ \citenamefont {Tokura}}]{yu2012skyrmion}%
  \BibitemOpen
  \bibfield  {author} {\bibinfo {author} {\bibfnamefont {X.}~\bibnamefont {Yu}}, \bibinfo {author} {\bibfnamefont {N.}~\bibnamefont {Kanazawa}}, \bibinfo {author} {\bibfnamefont {W.}~\bibnamefont {Zhang}}, \bibinfo {author} {\bibfnamefont {T.}~\bibnamefont {Nagai}}, \bibinfo {author} {\bibfnamefont {T.}~\bibnamefont {Hara}}, \bibinfo {author} {\bibfnamefont {K.}~\bibnamefont {Kimoto}}, \bibinfo {author} {\bibfnamefont {Y.}~\bibnamefont {Matsui}}, \bibinfo {author} {\bibfnamefont {Y.}~\bibnamefont {Onose}}, \ and\ \bibinfo {author} {\bibfnamefont {Y.}~\bibnamefont {Tokura}},\ }\href {\doibase 10.1038/ncomms1990} {\bibfield  {journal} {\bibinfo  {journal} {Nat Commun}\ }\textbf {\bibinfo {volume} {3}},\ \bibinfo {pages} {988} (\bibinfo {year} {2012})}\BibitemShut {NoStop}%
\bibitem [{\citenamefont {Yu}\ \emph {et~al.}(2017)\citenamefont {Yu}, \citenamefont {Morikawa}, \citenamefont {Tokunaga}, \citenamefont {Kubota}, \citenamefont {Kurumaji}, \citenamefont {Oike}, \citenamefont {Nakamura}, \citenamefont {Kagawa}, \citenamefont {Taguchi}, \citenamefont {Arima}, \citenamefont {Kawasaki},\ and\ \citenamefont {Tokura}}]{yu2017current}%
  \BibitemOpen
  \bibfield  {author} {\bibinfo {author} {\bibfnamefont {X.}~\bibnamefont {Yu}}, \bibinfo {author} {\bibfnamefont {D.}~\bibnamefont {Morikawa}}, \bibinfo {author} {\bibfnamefont {Y.}~\bibnamefont {Tokunaga}}, \bibinfo {author} {\bibfnamefont {M.}~\bibnamefont {Kubota}}, \bibinfo {author} {\bibfnamefont {T.}~\bibnamefont {Kurumaji}}, \bibinfo {author} {\bibfnamefont {H.}~\bibnamefont {Oike}}, \bibinfo {author} {\bibfnamefont {M.}~\bibnamefont {Nakamura}}, \bibinfo {author} {\bibfnamefont {F.}~\bibnamefont {Kagawa}}, \bibinfo {author} {\bibfnamefont {Y.}~\bibnamefont {Taguchi}}, \bibinfo {author} {\bibfnamefont {T.-h.}\ \bibnamefont {Arima}}, \bibinfo {author} {\bibfnamefont {M.}~\bibnamefont {Kawasaki}}, \ and\ \bibinfo {author} {\bibfnamefont {Y.}~\bibnamefont {Tokura}},\ }\href {\doibase 10.1002/adma.201606178} {\bibfield  {journal} {\bibinfo  {journal} {Advanced Materials}\ }\textbf {\bibinfo {volume} {29}},\ \bibinfo {pages} {1606178} (\bibinfo {year} {2017})}\BibitemShut {NoStop}%
\bibitem [{\citenamefont {Weiler}\ \emph {et~al.}(2017)\citenamefont {Weiler}, \citenamefont {Aqeel}, \citenamefont {Mostovoy}, \citenamefont {Leonov}, \citenamefont {Gepr{\"a}gs}, \citenamefont {Gross}, \citenamefont {Huebl}, \citenamefont {Palstra},\ and\ \citenamefont {Goennenwein}}]{weiler2017helimagnona}%
  \BibitemOpen
  \bibfield  {author} {\bibinfo {author} {\bibfnamefont {M.}~\bibnamefont {Weiler}}, \bibinfo {author} {\bibfnamefont {A.}~\bibnamefont {Aqeel}}, \bibinfo {author} {\bibfnamefont {M.}~\bibnamefont {Mostovoy}}, \bibinfo {author} {\bibfnamefont {A.}~\bibnamefont {Leonov}}, \bibinfo {author} {\bibfnamefont {S.}~\bibnamefont {Gepr{\"a}gs}}, \bibinfo {author} {\bibfnamefont {R.}~\bibnamefont {Gross}}, \bibinfo {author} {\bibfnamefont {H.}~\bibnamefont {Huebl}}, \bibinfo {author} {\bibfnamefont {T.~T.~M.}\ \bibnamefont {Palstra}}, \ and\ \bibinfo {author} {\bibfnamefont {S.~T.~B.}\ \bibnamefont {Goennenwein}},\ }\href {\doibase 10.1103/PhysRevLett.119.237204} {\bibfield  {journal} {\bibinfo  {journal} {Phys. Rev. Lett.}\ }\textbf {\bibinfo {volume} {119}},\ \bibinfo {pages} {237204} (\bibinfo {year} {2017})}\BibitemShut {NoStop}%
\bibitem [{\citenamefont {Mochizuki}(2012)}]{mochizuki2012spinwavea}%
  \BibitemOpen
  \bibfield  {author} {\bibinfo {author} {\bibfnamefont {M.}~\bibnamefont {Mochizuki}},\ }\href {\doibase 10.1103/PhysRevLett.108.017601} {\bibfield  {journal} {\bibinfo  {journal} {Phys. Rev. Lett.}\ }\textbf {\bibinfo {volume} {108}},\ \bibinfo {pages} {017601} (\bibinfo {year} {2012})}\BibitemShut {NoStop}%
\bibitem [{\citenamefont {Shibata}\ \emph {et~al.}(2015)\citenamefont {Shibata}, \citenamefont {Iwasaki}, \citenamefont {Kanazawa}, \citenamefont {Aizawa}, \citenamefont {Tanigaki}, \citenamefont {Shirai}, \citenamefont {Nakajima}, \citenamefont {Kubota}, \citenamefont {Kawasaki}, \citenamefont {Park}, \citenamefont {Shindo}, \citenamefont {Nagaosa},\ and\ \citenamefont {Tokura}}]{shibata2015largea}%
  \BibitemOpen
  \bibfield  {author} {\bibinfo {author} {\bibfnamefont {K.}~\bibnamefont {Shibata}}, \bibinfo {author} {\bibfnamefont {J.}~\bibnamefont {Iwasaki}}, \bibinfo {author} {\bibfnamefont {N.}~\bibnamefont {Kanazawa}}, \bibinfo {author} {\bibfnamefont {S.}~\bibnamefont {Aizawa}}, \bibinfo {author} {\bibfnamefont {T.}~\bibnamefont {Tanigaki}}, \bibinfo {author} {\bibfnamefont {M.}~\bibnamefont {Shirai}}, \bibinfo {author} {\bibfnamefont {T.}~\bibnamefont {Nakajima}}, \bibinfo {author} {\bibfnamefont {M.}~\bibnamefont {Kubota}}, \bibinfo {author} {\bibfnamefont {M.}~\bibnamefont {Kawasaki}}, \bibinfo {author} {\bibfnamefont {H.~S.}\ \bibnamefont {Park}}, \bibinfo {author} {\bibfnamefont {D.}~\bibnamefont {Shindo}}, \bibinfo {author} {\bibfnamefont {N.}~\bibnamefont {Nagaosa}}, \ and\ \bibinfo {author} {\bibfnamefont {Y.}~\bibnamefont {Tokura}},\ }\href {\doibase 10.1038/nnano.2015.113} {\bibfield  {journal} {\bibinfo  {journal} {Nature Nanotech}\ }\textbf {\bibinfo {volume} {10}},\ \bibinfo {pages} {589} (\bibinfo {year} {2015})}\BibitemShut {NoStop}%
\bibitem [{\citenamefont {Zhang}\ \emph {et~al.}(2021)\citenamefont {Zhang}, \citenamefont {Liu}, \citenamefont {Dong}, \citenamefont {Wu}, \citenamefont {Zhang}, \citenamefont {Wang}, \citenamefont {Lu}, \citenamefont {R{\"u}ckriegel}, \citenamefont {Wang}, \citenamefont {Duine}, \citenamefont {Yu}, \citenamefont {Luo}, \citenamefont {Shen},\ and\ \citenamefont {Zhang}}]{zhang2021straindrivena}%
  \BibitemOpen
  \bibfield  {author} {\bibinfo {author} {\bibfnamefont {Y.}~\bibnamefont {Zhang}}, \bibinfo {author} {\bibfnamefont {J.}~\bibnamefont {Liu}}, \bibinfo {author} {\bibfnamefont {Y.}~\bibnamefont {Dong}}, \bibinfo {author} {\bibfnamefont {S.}~\bibnamefont {Wu}}, \bibinfo {author} {\bibfnamefont {J.}~\bibnamefont {Zhang}}, \bibinfo {author} {\bibfnamefont {J.}~\bibnamefont {Wang}}, \bibinfo {author} {\bibfnamefont {J.}~\bibnamefont {Lu}}, \bibinfo {author} {\bibfnamefont {A.}~\bibnamefont {R{\"u}ckriegel}}, \bibinfo {author} {\bibfnamefont {H.}~\bibnamefont {Wang}}, \bibinfo {author} {\bibfnamefont {R.}~\bibnamefont {Duine}}, \bibinfo {author} {\bibfnamefont {H.}~\bibnamefont {Yu}}, \bibinfo {author} {\bibfnamefont {Z.}~\bibnamefont {Luo}}, \bibinfo {author} {\bibfnamefont {K.}~\bibnamefont {Shen}}, \ and\ \bibinfo {author} {\bibfnamefont {J.}~\bibnamefont {Zhang}},\ }\href {\doibase 10.1103/PhysRevLett.127.117204} {\bibfield  {journal} {\bibinfo  {journal} {Phys. Rev. Lett.}\ }\textbf {\bibinfo {volume} {127}},\ \bibinfo {pages} {117204} (\bibinfo {year} {2021})}\BibitemShut {NoStop}%
\bibitem [{\citenamefont {Lei}\ \emph {et~al.}(2013)\citenamefont {Lei}, \citenamefont {Devolder}, \citenamefont {Agnus}, \citenamefont {Aubert}, \citenamefont {Daniel}, \citenamefont {Kim}, \citenamefont {Zhao}, \citenamefont {Trypiniotis}, \citenamefont {Cowburn}, \citenamefont {Chappert}, \citenamefont {Ravelosona},\ and\ \citenamefont {Lecoeur}}]{lei2013straincontrolled}%
  \BibitemOpen
  \bibfield  {author} {\bibinfo {author} {\bibfnamefont {N.}~\bibnamefont {Lei}}, \bibinfo {author} {\bibfnamefont {T.}~\bibnamefont {Devolder}}, \bibinfo {author} {\bibfnamefont {G.}~\bibnamefont {Agnus}}, \bibinfo {author} {\bibfnamefont {P.}~\bibnamefont {Aubert}}, \bibinfo {author} {\bibfnamefont {L.}~\bibnamefont {Daniel}}, \bibinfo {author} {\bibfnamefont {J.-V.}\ \bibnamefont {Kim}}, \bibinfo {author} {\bibfnamefont {W.}~\bibnamefont {Zhao}}, \bibinfo {author} {\bibfnamefont {T.}~\bibnamefont {Trypiniotis}}, \bibinfo {author} {\bibfnamefont {R.~P.}\ \bibnamefont {Cowburn}}, \bibinfo {author} {\bibfnamefont {C.}~\bibnamefont {Chappert}}, \bibinfo {author} {\bibfnamefont {D.}~\bibnamefont {Ravelosona}}, \ and\ \bibinfo {author} {\bibfnamefont {P.}~\bibnamefont {Lecoeur}},\ }\href {\doibase 10.1038/ncomms2386} {\bibfield  {journal} {\bibinfo  {journal} {Nat Commun}\ }\textbf {\bibinfo {volume} {4}},\ \bibinfo {pages} {1378} (\bibinfo {year} {2013})}\BibitemShut {NoStop}%
\bibitem [{\citenamefont {Kong}\ \emph {et~al.}(2023)\citenamefont {Kong}, \citenamefont {Kov{\'a}cs}, \citenamefont {Charilaou}, \citenamefont {Zheng}, \citenamefont {Wang}, \citenamefont {Han},\ and\ \citenamefont {{Dunin-Borkowski}}}]{kong2023direct}%
  \BibitemOpen
  \bibfield  {author} {\bibinfo {author} {\bibfnamefont {D.}~\bibnamefont {Kong}}, \bibinfo {author} {\bibfnamefont {A.}~\bibnamefont {Kov{\'a}cs}}, \bibinfo {author} {\bibfnamefont {M.}~\bibnamefont {Charilaou}}, \bibinfo {author} {\bibfnamefont {F.}~\bibnamefont {Zheng}}, \bibinfo {author} {\bibfnamefont {L.}~\bibnamefont {Wang}}, \bibinfo {author} {\bibfnamefont {X.}~\bibnamefont {Han}}, \ and\ \bibinfo {author} {\bibfnamefont {R.~E.}\ \bibnamefont {{Dunin-Borkowski}}},\ }\href {\doibase 10.1038/s41467-023-39650-8} {\bibfield  {journal} {\bibinfo  {journal} {Nat Commun}\ }\textbf {\bibinfo {volume} {14}},\ \bibinfo {pages} {3963} (\bibinfo {year} {2023})}\BibitemShut {NoStop}%
\bibitem [{\citenamefont {Wang}(2019)}]{wang2019mechanical}%
  \BibitemOpen
  \bibfield  {author} {\bibinfo {author} {\bibfnamefont {J.}~\bibnamefont {Wang}},\ }\href {\doibase 10.1146/annurev-matsci-070218-010200} {\bibfield  {journal} {\bibinfo  {journal} {Annu. Rev. Mater. Res.}\ }\textbf {\bibinfo {volume} {49}},\ \bibinfo {pages} {361} (\bibinfo {year} {2019})}\BibitemShut {NoStop}%
\bibitem [{\citenamefont {Bukharaev}\ \emph {et~al.}(2018)\citenamefont {Bukharaev}, \citenamefont {Zvezdin}, \citenamefont {Pyatakov},\ and\ \citenamefont {Fetisov}}]{bukharaev2018straintronics}%
  \BibitemOpen
  \bibfield  {author} {\bibinfo {author} {\bibfnamefont {A.~A.}\ \bibnamefont {Bukharaev}}, \bibinfo {author} {\bibfnamefont {A.~K.}\ \bibnamefont {Zvezdin}}, \bibinfo {author} {\bibfnamefont {A.~P.}\ \bibnamefont {Pyatakov}}, \ and\ \bibinfo {author} {\bibfnamefont {Y.~K.}\ \bibnamefont {Fetisov}},\ }\href {\doibase 10.3367/UFNe.2018.01.038279} {\bibfield  {journal} {\bibinfo  {journal} {Phys.-Usp.}\ }\textbf {\bibinfo {volume} {61}},\ \bibinfo {pages} {1175} (\bibinfo {year} {2018})}\BibitemShut {NoStop}%
\bibitem [{\citenamefont {Gopman}\ \emph {et~al.}(2016)\citenamefont {Gopman}, \citenamefont {Dennis}, \citenamefont {Chen}, \citenamefont {Iunin}, \citenamefont {Finkel}, \citenamefont {Staruch},\ and\ \citenamefont {Shull}}]{gopman2016strainassisted}%
  \BibitemOpen
  \bibfield  {author} {\bibinfo {author} {\bibfnamefont {D.~B.}\ \bibnamefont {Gopman}}, \bibinfo {author} {\bibfnamefont {C.~L.}\ \bibnamefont {Dennis}}, \bibinfo {author} {\bibfnamefont {P.~J.}\ \bibnamefont {Chen}}, \bibinfo {author} {\bibfnamefont {Y.~L.}\ \bibnamefont {Iunin}}, \bibinfo {author} {\bibfnamefont {P.}~\bibnamefont {Finkel}}, \bibinfo {author} {\bibfnamefont {M.}~\bibnamefont {Staruch}}, \ and\ \bibinfo {author} {\bibfnamefont {R.~D.}\ \bibnamefont {Shull}},\ }\href {\doibase 10.1038/srep27774} {\bibfield  {journal} {\bibinfo  {journal} {Sci Rep}\ }\textbf {\bibinfo {volume} {6}},\ \bibinfo {pages} {27774} (\bibinfo {year} {2016})}\BibitemShut {NoStop}%
\bibitem [{\citenamefont {Sukhanov}\ \emph {et~al.}(2019)\citenamefont {Sukhanov}, \citenamefont {Vir}, \citenamefont {Heinemann}, \citenamefont {Nikitin}, \citenamefont {Kriegner}, \citenamefont {Borrmann}, \citenamefont {Shekhar}, \citenamefont {Felser},\ and\ \citenamefont {Inosov}}]{sukhanov2019gianta}%
  \BibitemOpen
  \bibfield  {author} {\bibinfo {author} {\bibfnamefont {A.~S.}\ \bibnamefont {Sukhanov}}, \bibinfo {author} {\bibfnamefont {P.}~\bibnamefont {Vir}}, \bibinfo {author} {\bibfnamefont {A.}~\bibnamefont {Heinemann}}, \bibinfo {author} {\bibfnamefont {S.~E.}\ \bibnamefont {Nikitin}}, \bibinfo {author} {\bibfnamefont {D.}~\bibnamefont {Kriegner}}, \bibinfo {author} {\bibfnamefont {H.}~\bibnamefont {Borrmann}}, \bibinfo {author} {\bibfnamefont {C.}~\bibnamefont {Shekhar}}, \bibinfo {author} {\bibfnamefont {C.}~\bibnamefont {Felser}}, \ and\ \bibinfo {author} {\bibfnamefont {D.~S.}\ \bibnamefont {Inosov}},\ }\href {\doibase 10.1103/PhysRevB.100.180403} {\bibfield  {journal} {\bibinfo  {journal} {Phys. Rev. B}\ }\textbf {\bibinfo {volume} {100}},\ \bibinfo {pages} {180403} (\bibinfo {year} {2019})}\BibitemShut {NoStop}%
\bibitem [{\citenamefont {Nii}\ \emph {et~al.}(2015)\citenamefont {Nii}, \citenamefont {Nakajima}, \citenamefont {Kikkawa}, \citenamefont {Yamasaki}, \citenamefont {Ohishi}, \citenamefont {Suzuki}, \citenamefont {Taguchi}, \citenamefont {Arima}, \citenamefont {Tokura},\ and\ \citenamefont {Iwasa}}]{nii2015uniaxiala}%
  \BibitemOpen
  \bibfield  {author} {\bibinfo {author} {\bibfnamefont {Y.}~\bibnamefont {Nii}}, \bibinfo {author} {\bibfnamefont {T.}~\bibnamefont {Nakajima}}, \bibinfo {author} {\bibfnamefont {A.}~\bibnamefont {Kikkawa}}, \bibinfo {author} {\bibfnamefont {Y.}~\bibnamefont {Yamasaki}}, \bibinfo {author} {\bibfnamefont {K.}~\bibnamefont {Ohishi}}, \bibinfo {author} {\bibfnamefont {J.}~\bibnamefont {Suzuki}}, \bibinfo {author} {\bibfnamefont {Y.}~\bibnamefont {Taguchi}}, \bibinfo {author} {\bibfnamefont {T.}~\bibnamefont {Arima}}, \bibinfo {author} {\bibfnamefont {Y.}~\bibnamefont {Tokura}}, \ and\ \bibinfo {author} {\bibfnamefont {Y.}~\bibnamefont {Iwasa}},\ }\href {\doibase 10.1038/ncomms9539} {\bibfield  {journal} {\bibinfo  {journal} {Nat Commun}\ }\textbf {\bibinfo {volume} {6}},\ \bibinfo {pages} {8539} (\bibinfo {year} {2015})}\BibitemShut {NoStop}%
\bibitem [{\citenamefont {Paterson}\ \emph {et~al.}(2020)\citenamefont {Paterson}, \citenamefont {Tereshchenko}, \citenamefont {Nakayama}, \citenamefont {Kousaka}, \citenamefont {Kishine}, \citenamefont {McVitie}, \citenamefont {Ovchinnikov}, \citenamefont {Proskurin},\ and\ \citenamefont {Togawa}}]{paterson2020tensile}%
  \BibitemOpen
  \bibfield  {author} {\bibinfo {author} {\bibfnamefont {G.~W.}\ \bibnamefont {Paterson}}, \bibinfo {author} {\bibfnamefont {A.~A.}\ \bibnamefont {Tereshchenko}}, \bibinfo {author} {\bibfnamefont {S.}~\bibnamefont {Nakayama}}, \bibinfo {author} {\bibfnamefont {Y.}~\bibnamefont {Kousaka}}, \bibinfo {author} {\bibfnamefont {J.}~\bibnamefont {Kishine}}, \bibinfo {author} {\bibfnamefont {S.}~\bibnamefont {McVitie}}, \bibinfo {author} {\bibfnamefont {A.~S.}\ \bibnamefont {Ovchinnikov}}, \bibinfo {author} {\bibfnamefont {I.}~\bibnamefont {Proskurin}}, \ and\ \bibinfo {author} {\bibfnamefont {Y.}~\bibnamefont {Togawa}},\ }\href {\doibase 10.1103/PhysRevB.101.184424} {\bibfield  {journal} {\bibinfo  {journal} {Phys. Rev. B}\ }\textbf {\bibinfo {volume} {101}},\ \bibinfo {pages} {184424} (\bibinfo {year} {2020})}\BibitemShut {NoStop}%
\bibitem [{\citenamefont {Plumer}\ and\ \citenamefont {Walker}(1982)}]{plumer1982magnetoelastic}%
  \BibitemOpen
  \bibfield  {author} {\bibinfo {author} {\bibfnamefont {M.~L.}\ \bibnamefont {Plumer}}\ and\ \bibinfo {author} {\bibfnamefont {M.~B.}\ \bibnamefont {Walker}},\ }\href {\doibase 10.1088/0022-3719/15/35/015} {\bibfield  {journal} {\bibinfo  {journal} {J. Phys. C: Solid State Phys.}\ }\textbf {\bibinfo {volume} {15}},\ \bibinfo {pages} {7181} (\bibinfo {year} {1982})}\BibitemShut {NoStop}%
\bibitem [{\citenamefont {Koretsune}\ \emph {et~al.}(2015)\citenamefont {Koretsune}, \citenamefont {Nagaosa},\ and\ \citenamefont {Arita}}]{koretsune2015controla}%
  \BibitemOpen
  \bibfield  {author} {\bibinfo {author} {\bibfnamefont {T.}~\bibnamefont {Koretsune}}, \bibinfo {author} {\bibfnamefont {N.}~\bibnamefont {Nagaosa}}, \ and\ \bibinfo {author} {\bibfnamefont {R.}~\bibnamefont {Arita}},\ }\href {\doibase 10.1038/srep13302} {\bibfield  {journal} {\bibinfo  {journal} {Sci Rep}\ }\textbf {\bibinfo {volume} {5}},\ \bibinfo {pages} {13302} (\bibinfo {year} {2015})}\BibitemShut {NoStop}%
\bibitem [{Sup()}]{Supplementary}%
  \BibitemOpen
  \href@noop {} {}\bibinfo {note} {See Supplemental Material at \url{URL_will_be_inserted_by_publisher}, which includes Refs. \cite{ishizuka2005Phase_,landauElectrodynamics_,dzyaloshinskii1965Theory_,leonov2016Chiral_,hals2017New_,rybakov2016New_,brown1963Thermal_, leliaert2017Adaptively_}, for further details on the experimental methods, as well as a comprehensive theoretical discussion of the framework (including analytical calculations) and numerical simulations.}\BibitemShut {Stop}%
\bibitem [{\citenamefont {Ishizuka}\ and\ \citenamefont {Allman}(2005)}]{ishizuka2005Phase_}%
  \BibitemOpen
  \bibfield  {author} {\bibinfo {author} {\bibfnamefont {K.}~\bibnamefont {Ishizuka}}\ and\ \bibinfo {author} {\bibfnamefont {B.}~\bibnamefont {Allman}},\ }\href {\doibase 10.1017/S1551929500051592} {\bibfield  {journal} {\bibinfo  {journal} {Microscopy Today}\ }\textbf {\bibinfo {volume} {13}},\ \bibinfo {pages} {22} (\bibinfo {year} {2005})}\BibitemShut {NoStop}%
\bibitem [{\citenamefont {Landau}\ \emph {et~al.}()\citenamefont {Landau}, \citenamefont {Pitaevskii},\ and\ \citenamefont {Lifshitz}}]{landauElectrodynamics_}%
  \BibitemOpen
  \bibfield  {author} {\bibinfo {author} {\bibfnamefont {L.~D.}\ \bibnamefont {Landau}}, \bibinfo {author} {\bibfnamefont {L.~P.}\ \bibnamefont {Pitaevskii}}, \ and\ \bibinfo {author} {\bibfnamefont {E.~M.}\ \bibnamefont {Lifshitz}},\ }\href@noop {} {\emph {\bibinfo {title} {Electrodynamics of {{Continuous Media}}, {{Course}} of {{Theoretical Physics}} ({{Pergamon}}, {{Oxford}}, 1984)}}},\ Vol.~\bibinfo {volume} {8}\BibitemShut {NoStop}%
\bibitem [{\citenamefont {Dzyaloshinskii}(1965)}]{dzyaloshinskii1965Theory_}%
  \BibitemOpen
  \bibfield  {author} {\bibinfo {author} {\bibfnamefont {I.~E.}\ \bibnamefont {Dzyaloshinskii}},\ }\href@noop {} {\bibfield  {journal} {\bibinfo  {journal} {III. Sov. Phys. JETP}\ }\textbf {\bibinfo {volume} {20}},\ \bibinfo {pages} {665} (\bibinfo {year} {1965})}\BibitemShut {NoStop}%
\bibitem [{\citenamefont {Leonov}\ \emph {et~al.}(2016)\citenamefont {Leonov}, \citenamefont {Togawa}, \citenamefont {Monchesky}, \citenamefont {Bogdanov}, \citenamefont {Kishine}, \citenamefont {Kousaka}, \citenamefont {Miyagawa}, \citenamefont {Koyama}, \citenamefont {Akimitsu}, \citenamefont {Koyama}, \citenamefont {Harada}, \citenamefont {Mori}, \citenamefont {McGrouther}, \citenamefont {Lamb}, \citenamefont {Krajnak}, \citenamefont {McVitie}, \citenamefont {Stamps},\ and\ \citenamefont {Inoue}}]{leonov2016Chiral_}%
  \BibitemOpen
  \bibfield  {author} {\bibinfo {author} {\bibfnamefont {A.~O.}\ \bibnamefont {Leonov}}, \bibinfo {author} {\bibfnamefont {Y.}~\bibnamefont {Togawa}}, \bibinfo {author} {\bibfnamefont {T.~L.}\ \bibnamefont {Monchesky}}, \bibinfo {author} {\bibfnamefont {A.~N.}\ \bibnamefont {Bogdanov}}, \bibinfo {author} {\bibfnamefont {J.}~\bibnamefont {Kishine}}, \bibinfo {author} {\bibfnamefont {Y.}~\bibnamefont {Kousaka}}, \bibinfo {author} {\bibfnamefont {M.}~\bibnamefont {Miyagawa}}, \bibinfo {author} {\bibfnamefont {T.}~\bibnamefont {Koyama}}, \bibinfo {author} {\bibfnamefont {J.}~\bibnamefont {Akimitsu}}, \bibinfo {author} {\bibfnamefont {{\relax Ts}.}~\bibnamefont {Koyama}}, \bibinfo {author} {\bibfnamefont {K.}~\bibnamefont {Harada}}, \bibinfo {author} {\bibfnamefont {S.}~\bibnamefont {Mori}}, \bibinfo {author} {\bibfnamefont {D.}~\bibnamefont {McGrouther}}, \bibinfo {author} {\bibfnamefont {R.}~\bibnamefont {Lamb}}, \bibinfo {author} {\bibfnamefont {M.}~\bibnamefont {Krajnak}}, \bibinfo {author} {\bibfnamefont {S.}~\bibnamefont {McVitie}}, \bibinfo {author} {\bibfnamefont {R.~L.}\ \bibnamefont {Stamps}}, \ and\ \bibinfo {author} {\bibfnamefont {K.}~\bibnamefont {Inoue}},\ }\href {\doibase 10.1103/PhysRevLett.117.087202} {\bibfield  {journal} {\bibinfo  {journal} {Physical Review Letters}\ }\textbf {\bibinfo {volume} {117}},\ \bibinfo {pages} {087202} (\bibinfo {year} {2016})}\BibitemShut {NoStop}%
\bibitem [{\citenamefont {Hals}\ and\ \citenamefont {{Everschor-Sitte}}(2017)}]{hals2017New_}%
  \BibitemOpen
  \bibfield  {author} {\bibinfo {author} {\bibfnamefont {K.~M.~D.}\ \bibnamefont {Hals}}\ and\ \bibinfo {author} {\bibfnamefont {K.}~\bibnamefont {{Everschor-Sitte}}},\ }\href {\doibase 10.1103/PhysRevLett.119.127203} {\bibfield  {journal} {\bibinfo  {journal} {Physical Review Letters}\ }\textbf {\bibinfo {volume} {119}},\ \bibinfo {pages} {127203} (\bibinfo {year} {2017})}\BibitemShut {NoStop}%
\bibitem [{\citenamefont {Rybakov}\ \emph {et~al.}(2016)\citenamefont {Rybakov}, \citenamefont {Borisov}, \citenamefont {Bl{\"u}gel},\ and\ \citenamefont {Kiselev}}]{rybakov2016New_}%
  \BibitemOpen
  \bibfield  {author} {\bibinfo {author} {\bibfnamefont {F.~N.}\ \bibnamefont {Rybakov}}, \bibinfo {author} {\bibfnamefont {A.~B.}\ \bibnamefont {Borisov}}, \bibinfo {author} {\bibfnamefont {S.}~\bibnamefont {Bl{\"u}gel}}, \ and\ \bibinfo {author} {\bibfnamefont {N.~S.}\ \bibnamefont {Kiselev}},\ }\href {\doibase 10.1088/1367-2630/18/4/045002} {\bibfield  {journal} {\bibinfo  {journal} {New Journal of Physics}\ }\textbf {\bibinfo {volume} {18}},\ \bibinfo {pages} {045002} (\bibinfo {year} {2016})}\BibitemShut {NoStop}%
\bibitem [{\citenamefont {Brown}(1963)}]{brown1963Thermal_}%
  \BibitemOpen
  \bibfield  {author} {\bibinfo {author} {\bibfnamefont {W.~F.}\ \bibnamefont {Brown}},\ }\href {\doibase 10.1103/PhysRev.130.1677} {\bibfield  {journal} {\bibinfo  {journal} {Physical Review}\ }\textbf {\bibinfo {volume} {130}},\ \bibinfo {pages} {1677} (\bibinfo {year} {1963})}\BibitemShut {NoStop}%
\bibitem [{\citenamefont {Leliaert}\ \emph {et~al.}(2017)\citenamefont {Leliaert}, \citenamefont {Mulkers}, \citenamefont {De~Clercq}, \citenamefont {Coene}, \citenamefont {Dvornik},\ and\ \citenamefont {Van~Waeyenberge}}]{leliaert2017Adaptively_}%
  \BibitemOpen
  \bibfield  {author} {\bibinfo {author} {\bibfnamefont {J.}~\bibnamefont {Leliaert}}, \bibinfo {author} {\bibfnamefont {J.}~\bibnamefont {Mulkers}}, \bibinfo {author} {\bibfnamefont {J.}~\bibnamefont {De~Clercq}}, \bibinfo {author} {\bibfnamefont {A.}~\bibnamefont {Coene}}, \bibinfo {author} {\bibfnamefont {M.}~\bibnamefont {Dvornik}}, \ and\ \bibinfo {author} {\bibfnamefont {B.}~\bibnamefont {Van~Waeyenberge}},\ }\href {\doibase 10.1063/1.5003957} {\bibfield  {journal} {\bibinfo  {journal} {AIP Advances}\ }\textbf {\bibinfo {volume} {7}},\ \bibinfo {pages} {125010} (\bibinfo {year} {2017})}\BibitemShut {NoStop}%
\bibitem [{\citenamefont {Yu}\ \emph {et~al.}(2018)\citenamefont {Yu}, \citenamefont {Koshibae}, \citenamefont {Tokunaga}, \citenamefont {Shibata}, \citenamefont {Taguchi}, \citenamefont {Nagaosa},\ and\ \citenamefont {Tokura}}]{yu2018transformation}%
  \BibitemOpen
  \bibfield  {author} {\bibinfo {author} {\bibfnamefont {X.~Z.}\ \bibnamefont {Yu}}, \bibinfo {author} {\bibfnamefont {W.}~\bibnamefont {Koshibae}}, \bibinfo {author} {\bibfnamefont {Y.}~\bibnamefont {Tokunaga}}, \bibinfo {author} {\bibfnamefont {K.}~\bibnamefont {Shibata}}, \bibinfo {author} {\bibfnamefont {Y.}~\bibnamefont {Taguchi}}, \bibinfo {author} {\bibfnamefont {N.}~\bibnamefont {Nagaosa}}, \ and\ \bibinfo {author} {\bibfnamefont {Y.}~\bibnamefont {Tokura}},\ }\href {\doibase 10.1038/s41586-018-0745-3} {\bibfield  {journal} {\bibinfo  {journal} {Nature}\ }\textbf {\bibinfo {volume} {564}},\ \bibinfo {pages} {95} (\bibinfo {year} {2018})}\BibitemShut {NoStop}%
\bibitem [{\citenamefont {Nose}\ \emph {et~al.}(2010)\citenamefont {Nose}, \citenamefont {Miyanishi}, \citenamefont {Aizawa}, \citenamefont {Ito},\ and\ \citenamefont {Honma}}]{nose2010rotationala}%
  \BibitemOpen
  \bibfield  {author} {\bibinfo {author} {\bibfnamefont {T.}~\bibnamefont {Nose}}, \bibinfo {author} {\bibfnamefont {T.}~\bibnamefont {Miyanishi}}, \bibinfo {author} {\bibfnamefont {Y.}~\bibnamefont {Aizawa}}, \bibinfo {author} {\bibfnamefont {R.}~\bibnamefont {Ito}}, \ and\ \bibinfo {author} {\bibfnamefont {M.}~\bibnamefont {Honma}},\ }\href {\doibase 10.1143/JJAP.49.051701} {\bibfield  {journal} {\bibinfo  {journal} {Jpn. J. Appl. Phys.}\ }\textbf {\bibinfo {volume} {49}},\ \bibinfo {pages} {051701} (\bibinfo {year} {2010})}\BibitemShut {NoStop}%
\end{thebibliography}
\end{document}